# Transparent conductive oxides as a material platform for photonic neural networks

J. Gosciniak[1*] and J.B. Khurgin[2]

[1]ENSEMBLE3 sp. z o.o., Wolczynska 133, 01-919 Warsaw, Poland
[2]Electrical and Computer Engineering Department, Johns Hopkins University, Baltimore, Maryland 21218, USA
Corresponding authors: * jeckug10@yahoo.com.sg

**Abstract**
Photonics integrated circuits have a huge potential to serve as a framework for a new class of information processing machines and can enable ultrafast artificial neural networks. They can overcome the existing speed and power limits of the electronic processing elements and provide additional benefits of photonics such as high-bandwidth, sub-nanosecond latencies and low-energy interconnect credentials leading to a new paradigm called neuromorphic photonics. The main obstacle to realize such a task is a lack of proper material platform that imposes serious requirements on the architecture of the network. Here we suggest and justify that transparent conductive oxides can be an excellent candidate for such a task as they provide a nonlinearity and bistability under both optical and electrical inputs.

**Introduction**
With a growing demand for fast and efficient signal processing the traditional computation techniques becomes very inefficient due to fundamental speed and energy limitations. Traditional computers are built based on the so-called von Neumann architecture with two separated units, memory and processing, that operate in a sequential way [**1**]. In comparison, a brain can process all of the signals in a parallel way what provides huge benefits in terms of speed and energy efficiency [**2, 3**]. One of the first solution to overcome those limitations rely on simulating a neural network in software on traditional computers but the problem still exist as the efficiency is mostly limited by the transfer of data between memory and processor [**1**]. The neuromorphic computing promises to overcome those limitations by developing a hardware that can mimics biological brains *i.e.*, developing the artificial neurons and synapses and combining it into network [**4**].
Up to now, most of the efforts related to a neuromorphic computing referred to neuromorphic electronics which, however, suffer from a significant bandwidth limitation [**4, 5**]. On a contrary, neuromorphic photonics is considered as a very promising technology platform to surpass conventional electronics in many aspects [**6-10**]. It exploits photons to carry out matrix multiplication with almost zero power consumption and with almost unlimited bandwidth being simultaneously compatible with standard CMOS technology. They can achieve a fully parallelism by busing multiple signals on a single waveguide at the speed of light. Furthermore, the optical weights offer unique advantages in the latency of the computation. As a result, in recent years a lot of efforts were put into photonics neural networks (PNN) as it claims orders of magnitude improvements in compute efficiencies against electronic counterparts [**11-13**]. However, to realize such high demanding tasks a proper material platform and architecture approach is still missing.
Recently, we have proposed transparent conductive oxides (TCOs) integrated on the waveguide platform as a valuable candidate for such a task as it provides nonlinearity and bistability under an optical power coupled to the waveguide or an applied voltage [**14**]. Such a system exhibits two stable states that depends on the history of the system thus can act as optical analogue of memristor or thyristor.

**Material platform and system arrangement**

Up to now, most of the efforts on material platform for a neuromorphic computing focused on magnetic alloys, metal oxides, phase change materials (PCM), ferroelectrics, 2D van der Waals materials or organic materials [**4**]. However, they suffer either from a slow phase transition and thus slow switching time (PCM) [**15-17**], very high operation voltage related to the high coercive field (ferroelectrics) [**18, 19**], high power operation (metal oxides for valence change memory) [**20,21**], very poor device-to-device stability resulting in deviation of the switching voltage and currents that finally provides to failure (electrochemical metallization cells) [**22, 23**], complex fabrication (2D van der Waals heterostructures) [**24, 25**] or low speed and incompatibility with standard CMOS technology (organic materials) [**26, 27**].

Transparent conductive oxides that belong to the epsilon-near-zero (ENZ) materials proved to be an excellent material for electro-optic modulation and other optoelectronic applications due to the large permittivity tunability under an applied voltage or a light illumination [**28-35**]. They are characterized by low optical losses, fast switching time and low switching voltage. Furthermore, they are CMOS-compatible and can be mass-produced with standard fabrication methods [**28-35**].

The generalized Drude formula for non-parabolic bands of TCOs can be expressed:

$$\varepsilon(\omega) = \varepsilon_\infty - \frac{N_c q^2}{\varepsilon_0 (\omega^2 + i\omega\gamma) m^*(E)}$$

where $\varepsilon_\infty$ is the permittivity due to the bound electrons, $N_c$ is the carrier density, $e$ is the electron charge, $\omega$ is the working frequency, $m^*(E)$ is the energy-dependent effective mass, and $\gamma$ is the scattering rate.

Here, with TCO placed in a slot waveguide (plasmonic or dielectric) (**Fig. 1a**), positive feedback between absorption and propagating power, leading to bistability, can be established. The interaction of light with TCO material can be highly enhanced by plasmonics that provides an additional field confinement into TCO. Apart from guiding a mode, plasmonic metal can also serve as the electrical contact or/and the heater depending on the requirements, thus, providing a dual functionality [**15, 16, 28,**].

The underlying mechanism is the nonlinear dependence of dielectric constant on power due to absorption of light. When energy of the incoming photons is above absorption edge, more carriers are created in the band that increases plasma frequency and reduces dielectric constant (**Fig. 1b**) but when the photon energy is below interband absorption edge, the effect is opposite as the free carrier absorption causes carriers in the non-parabolic band to heat up, increasing the effective mass and, in a consequence, a dielectric constant. Change of dielectric constant causes redistribution of the electric field in the slot mode - as absolute value of dielectric constant decreases the field tends to concentrate in the slot filled with TCO and absorption arises (**Fig. 1c**). By judiciously choosing the operational wavelength for a given doping level, the positive feedback appears in which absorption increases. Thus, a hysteresis region is established showing the mode attenuation as a function of the input optical power coupled to a waveguide (**Fig. 2**).

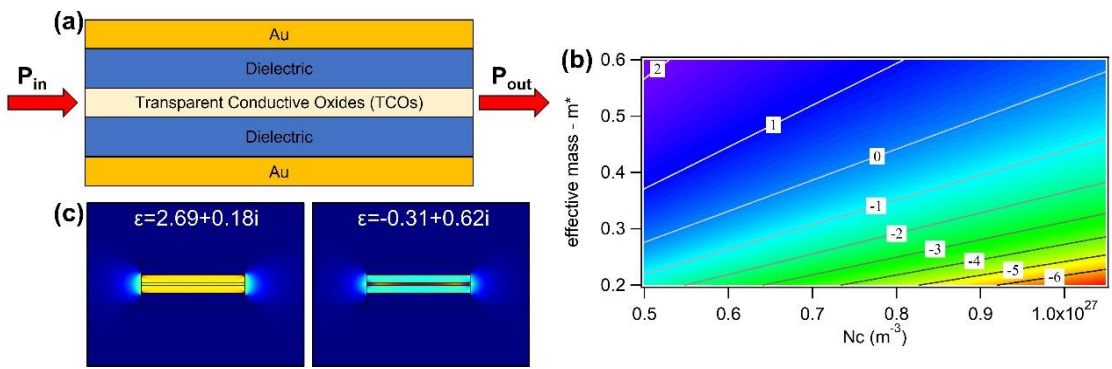

**Figure 1.** (a) Geometry of the proposed bistable device. (b) Dispersion of real part of dielectric permittivity as a function of effective mass and carrier concentration. (c) Electric field distribution in the state of low and high absorption regime.

When the light is off, the hot carriers (electrons) cool down in sub-picosecond time scale. For a low value of power coupled to the TCO, the system is stable and shows only one solution. However, for an increased value of power provided to a system, the three solutions appear. In this case, one of the solutions is unstable due to positive feedback. Starting from a low optical power, a gradual increases of power leads to gradual increases of effective mass (lower plasma frequency) and, in consequence, loss. Thus, only one stable solution exists. Further increases of power causes the second stable solution appears that corresponds to the critical (switching) value of power. When the power continues to increase, above the "switching power", the third unstable (middle) solution appears due to positive feedback. For a further increases of the optical power, the effective mass arises and TCO changes a sign from negative to positive. After this point, further increases of power causes only one stable solution [**14**].

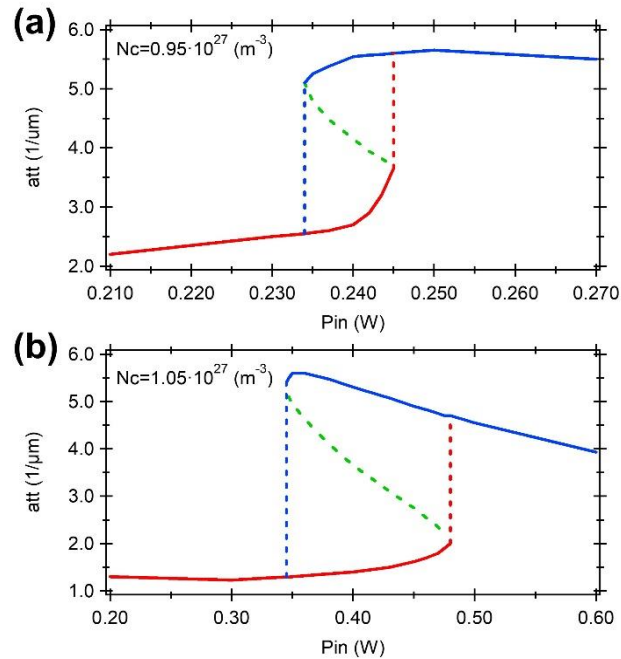

**Figure 2.** (a, b) Absorptive loss as a function of the propagating power exhibiting hysteresis manifesting all-optical bistability for different carrier concentration of ITO – (a) *Nc*=0.95·$10^{27}$ m$^{-3}$ and (b) *Nc*=1.05·$10^{27}$ m$^{-3}$.

In this way, for a certain values of input power, the bistability appears.

**Bistability in artificial neural networks**
Optical bistability refers to a nonlinear effect with feedback where network possesses two possible stable states for a single input [**36-42**]. Takin into account a computational point of view, the bistable systems are of particular interest because they can be used as memory units [**14, 41-42**]. In such systems, the ratio of output to input power ($P_{out}$ versus $P_{in}$) exhibit a hysteresis loop. By switching between the states, the system is able to represent, store, or retrace one bit of information. In a bistable network both states, low and high, are stable in the presence of small transient input pulses. Even if these transient input pulses modify the state on the network, the state returns to its initial state after the input pulse terminates. However, large input pulses or many low input pulses can induce transitions from one level to the other. Such nonlinear systems have a memory of their past state even though material does not have a memory itself [**41, 42, 43**].

Such bistable, nonlinear devices can serve as building blocks for neuromorphic computing architectures and information processing [17, 39, 41].

The concepts of machine learning and deep neural networks (DNNs) realized for artificial intelligence (AI) systems require a huge power budget to run deep learning algorithms thus the main goal of neuromorphic computing technologies is to reduce power consumption and improve the compute efficiency through the mimicking the computational capabilities of the brain [9, 11, 12, 43, 44]. Compared to conventional computation that is based on von Neumann architecture with the separated processing and memory units that is responsible for a large part of the power consumption [1], in the brain-inspired neuromorphic computing, the computation is performed in-place in the memory itself, thus it enables in-memory computing, what highly improve a power efficiency [17, 41, 43]. Furthermore, the brain-like computation is parallel and distributed where many simple summation nodes, neurons, replace the single central processing unit (CPU) of conventional computers and with computation results stored in the connection matrix [7].

The standard model of PNN comprise multiple neurons that can process input data signals based on the multilayer perceptron model. The perceptron is the basic building block of neural networks that exhibit both nonvolatile and volatile tunability [6, 7, 48].

**Model of perceptron**

Perceptron belongs to a first generation of artificial neural networks (ANNs) [13], that produces a binary output, 0 or 1, by a simple thresholding of the weighted synaptic input [6, 7]. A general model of perceptron is shown in **Fig. 3**. The input signals $x_i$ transmit the information from many presynaptic neurons to the postsynaptic neuron through weights $w_i$, which corresponds to the strength of the synapse. In the postsynaptic neuron, all the signals are summated in the cell body of the neuron and when the summated signal exceeds some threshold value, the signal is activated or if it is below some threshold value, the signal is deactivated. In general, two states are defined – one for activated signal that exceed some threshold value and other for deactivated signal that is below some threshold value. In such a way, the system keeps a memory about a given state of the neuron. Here, the bias $b$ represents an extra variable that remains in the system even if the rest of the inputs are absent. Sometimes, it can represent the random activation/firing of neurons that take place very often in our brains or can be associated with any other feature of the neuron.

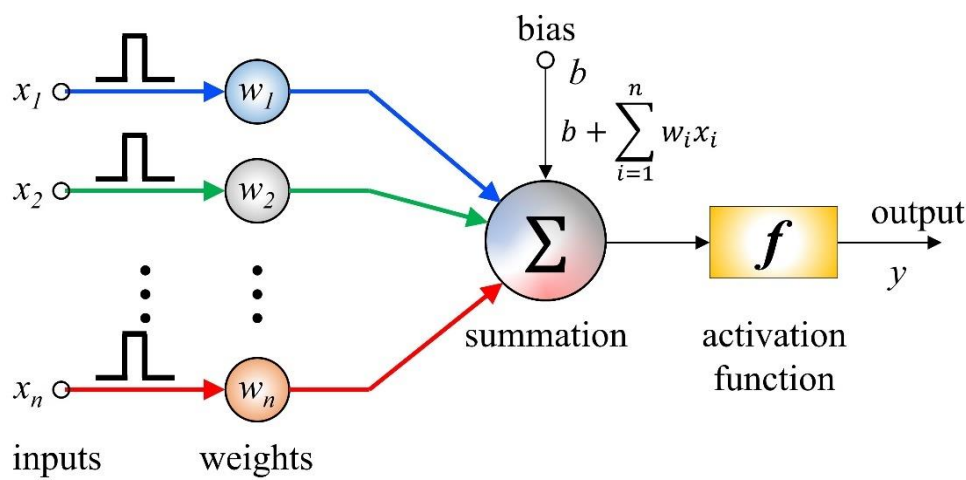

**Figure 3.** Schematic diagram of a perceptron with shown the fundamental operations in an artificial neuron: the weighted addition (or synaptic operation), summation and the threshold function.

From, the mathematical point of view, the operation principal can be described using the basic McCulloch-Pitts neuron model [2, 7, 46]. The output is expressed by $y=f(\sum w_i x_i + b)$, where $f$ is an activation (nonlinear) function, $x_i$ is the i-th element of the input vector $x$, $w_i$ is the weight factor for the input value $x_i$, and $b$ is a bias. Here, the linear term $\sum w_i x_i$ is called the weighted addition and corresponds to matrix-vector multiplications that consists of several vector-vector dot products $x_i w_i$. Since the operations performed by a perceptron are a collection of multiplication operations with a summation, the performance of a system is defined by multiply-accumulate (MAC) operations.

It should be mentioned that two types of summation can take place in biological neuron: spatial and temporal (**Fig. 6b, c**) [**2, 3**]. Spatial summation is about adding together of postsynaptic potentials generated simultaneously at many different synapses in a dendrite. Temporal summation is the adding together of postsynaptic potentials generated at the same synapse if they occur very fast, one after another, in the range of a few milliseconds. The perceptron model represents the spatial summation. However, the proposed arrangement can be implemented to serve for simultaneously for a temporal summation.

As seen from above, the overall process described by perceptron can be divided on three main parts: linear operation related with weighting, summation and nonlinear operation related with an activation function. All of them are very essential for the overall performance of the network, thus each part should be optimized in terms of performances. Furthermore, it will be highly desired if the system operates on a one material platform.

**All-optical switching**

One of the main advantages of the proposed geometry is that the same structure can be implemented simultaneously for linear operations (weighting), summation and non-linear operations. The linear operation requires a linear dependance between an output power and input power or applied voltage what take place much below a bistability point. The output signal depends linearly on the weight $w_i$ i.e., the length of TCO layer, its carrier concentration and input power. The output signal from each weight i.e., the vector-vector dot product $x_i w_i$, is then summed up and corresponds to matrix-vector multiplication, i.e., the weighted addition, $\sum w_i x_i$. Depending on the value of weighted addition the system can operate in low-loss or high-loss regime, thus in two possible logical states, "0" or "1", and can serve as the main storage unit.

Linear operations and weighting in photonic neural networks offer a lot of benefits in terms of speed, latency, and operation bandwidth [**4, 10, 47**]. The modulation speed exceeding 100 GHz can be achieved in some TCO-based modulators thanks to extremally fast carrier accumulation time through the electrical signal [**33**]. Furthermore, the performances can be further improved through all-optical modulation where the optical signal influences rather the effective mass of the TCO material than carrier concentration. Under an illumination of light, the effective mass of electrons in TCO changes what effect the plasma frequency, and in a consequence, the material permittivity. Furthermore, the optical modulation does not require E-O conversion. The speed in optical domain is limited only by the electron thermalization time in TCO that is much below picoseconds. Thus, a device can operate at speed exceeding hundreds of GHz.

As showed above, the system provides a lot of flexibility in terms of preferred operation schema. To optimize the system, it can operate in dual-mode operation, both electrical and optical, thus it can bring a lot of freedom in terms of design.

**Spatial summation**

One of the main obstacles in realization of the all-optical bistable devices is the large power requirements of the device which need at least millijoules per square centimeter to operate [**28**]. The problem can be solved either by low-loss, high-mobility conductive oxides with sharp resonances near the ENZ regime or by implementing the proposed bistable device as a memory unit that adds up many signals delivered to it.

Once the optical signals from all interconnects are weighted and then combined in one component, the summation and then a non-linear operation take place (**Fig. 4a**).

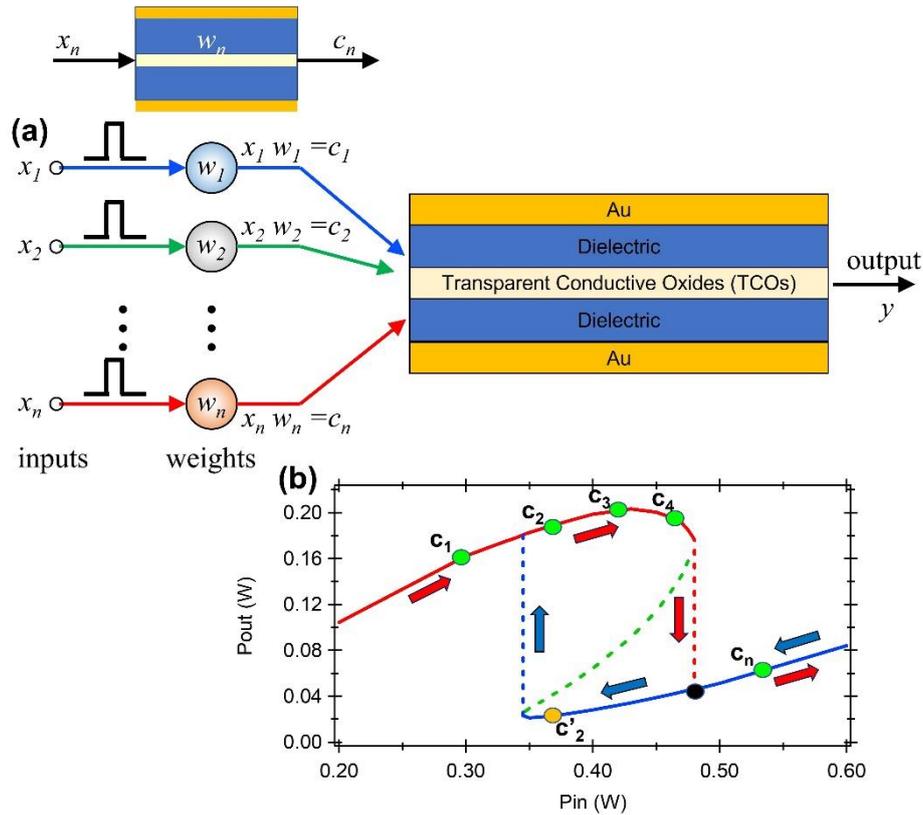

**Figure 4.** (a) Schematic diagram of a perceptron with a spatial summation and in a proposed arrangement with (b) operation principles.

The non-linear operation mimics the firing feature of biological neurons as it set a threshold from which activated and deactivated behaviors in artificial neurons are defined [**2, 3, 8, 48**].

Let's assume that at the beginning, the optical signal at the device shows certain value, for example defined by point $c_1$ or $c_2$. Such a situation is very common in the brain as neurons fire randomly so there is always some level of signal in the network. When signals are activated for a short period of time they combine in a device. When the overall signal, after summation, reaches point $c_4$ (**Fig. 4b**), and when the signals are off, the network returns to its initial state defined by point $c_2$. However, when the overall signal after summation exceeds a threshold value, here defined by a black dot in **Fig. 4b**, and when the signals are off, now the network follows the blue arrow and move to point $c'_2$ that is defined by lower output power. In such a way, a network keeps a memory about the processing that took place in the network. Similarly to the brain, enough pulses must be activated simultaneously to cause significant spatial summation of pulses. It is called cooperativity because coactive pulses must cooperate to produce enough signal to cause a transition from high to low transmission state [**3**].

Depending on the initial carrier concentration in the TCO, a device can work in different regime of input powers and under different range of attenuations.

Similarly to our previous paper [**14**], the calculations were performed for Au thickness of 50 nm, while the thickness of $Al_2O_3$ and ITO was taken at 50 nm and 10 nm, respectively. The width of the multilayer stack was taken at $w$=400 nm. The complex refractive index of Au at 1550 nm was assumed to be $n_{Au}$=0.596+10.923$i$ while the refractive index of $Al_2O_3$ $n_{Al2O3}$=1.621+0.00008$i$. The calculations were performed for the Fermi Energy $E_F$=4.5 eV, while a thermalization time was taken at $\tau$=500 ns. As observed, a thermalization time was much higher compared to our previous work ($\tau$=100 ns) what allowed to significantly reduce a power consumption.

For lower carrier concentration, the bistability range is limited to only of 10 mW input power, from 235 mW to 245 mW (**Fig. 5**), and in this range, the mode attenuation changes from around 2.6 $\mu m^{-1}$ to around 5.6 $\mu m^{-1}$ (**Fig. 2**). However, for higher carrier concentration, the bistability is present in much wider power range of 135 mW, from 345 mW to 480 mW (**Fig. 5**). In this power range, the mode attenuation changes from around 1.4 $\mu m^{-1}$ to around 5.6 $\mu m^{-1}$ (**Fig. 2**).

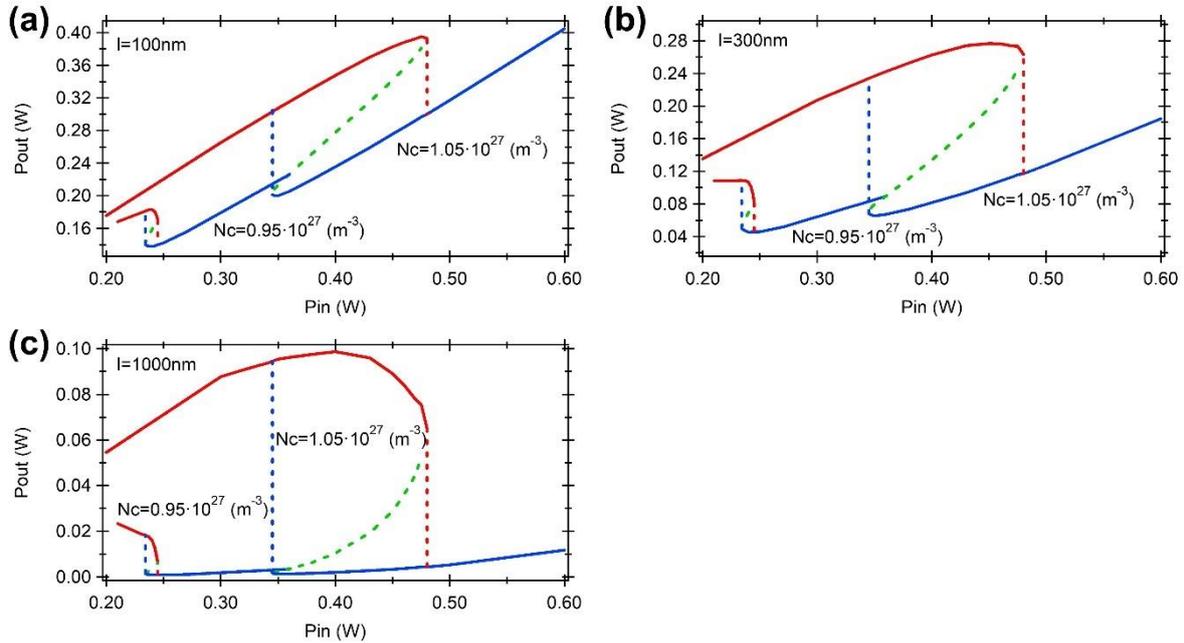

**Figure 5.** Input–output characteristics of the proposed bistable device of (a) 100 nm, (b) 300 nm and (c) 1000 nm length for different carrier concentration of 0.95·$10^{27}$ $m^{-3}$ and 1.05·$10^{27}$ $m^{-3}$.

Apart from the carrier concentration the system performances can be modified by the length of the TCO layer (**Fig. 5**). Mostly, it can influence the level of output power and a bistability range, however, it does not influence the input power range that defines a bistability. For short length of a device of 100nm, the output power difference is around 120 mW (from 230 mW to 350 mW) for an input power of 400 mW (**Fig. 5a**) and arise to 180 mW (from 80 mW to 260 mW) for 300 nm long device (**Fig. 5b**) and drops again to 98 mW (from 2 mW to 100 mW) for 1000 nm long devices (**Fig. 5c**). In the above cases, the input power was kept at 400 mW. For lower carrier concentration the output power difference is much smaller and range from 180 mW to 140 mW, 110 mW to 45 mW and from 18 mW to 1 mW for 100 nm, 300 nm, and 1000 nm long devices, respectively. The input power was kept constant at 235 mW.

Depending on an initial carrier concentration in TCO (here ITO) and length of the device, the system can work in different input powers range and under different range of bistability. For a low carrier concentration of 0.95·$10^{27}$ ($m^{-3}$) the output power changes from 138 mW to 183 mW for 100 nm long device, from 45 mW to 107 mW for 300 nm long device, and from 1 mW to 18 mW for 1000 nm long device, respectively. The values above refer to an input power of 240 mW.

The presented above schema can serve as a building block of the spatial architecture that represents the second generation of neural networks – deep neural networks (DNNs) where neurons (processing units) are stacked in processing layers, with adjacent layers interconnected via adjustable weights [**11, 12, 44**]. The output of a neuron is a weighted synaptic input that is transformed using a nonlinear activation function.

**Temporal summation**

As it was previously mentioned, apart from the spatial summation, the second type of summation can play an essential role in a brain processing – temporal summation [**2, 3**]. It refers to the process in which the consecutive synaptic potentials at the same site are added together in the postsynaptic cell [**2, 3**]. In this type of schema, the summation is controlled by the membrane time constant that helps determine the time course of the synaptic potential. Thus, neurons with a large membrane time constant have a greater capacity for temporal summation compared to neurons with a shorter time constant. The longer time constant means there is greater probability that two consecutive pulses from a presynaptic neuron will summate to bring the cell membrane to its threshold for an action potential. For a longer time constant, the first presynaptic potential does not fully decay by the time the second presynaptic potential is triggered. Therefore, the depolarizing effects of both potentials are additive what brings the membrane potential above the threshold and trigger an action potential [**2, 3, 48**]. In comparison, for a shorter time constant the first presynaptic potential decays to the resting potential before the second presynaptic potential is triggered. In consequence, the second presynaptic potential alone does not cause enough depolarization to trigger an action potential [**2**].

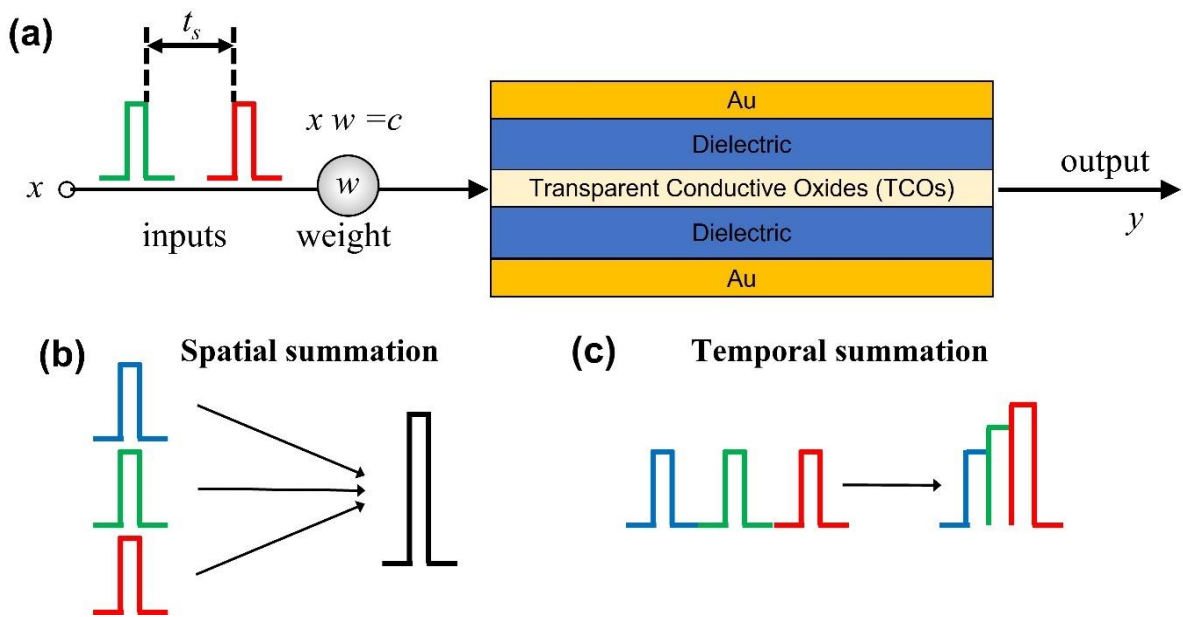

**Figure 6.** (a) Schematic diagram of a perceptron with a temporal summation in a proposed arrangement. (b, c) Comparison of (b) spatial summation and (c) temporal summation of three pulses in a proposed device.

The temporal summation can be implemented to the proposed structure (**Fig. 6**) where the membrane time constant is replaced by a thermalization time of electrons in the conduction band of TCO, $\tau_{th}$ [**14, 28, 34**]. Furthermore, it has not to be limited to only two pulses, but multiple number of pulses can be considered as well. For a separation between two consequent pulses $t_s$ shorter than a thermalization time $\tau_{th}$, $t_s<\tau_{th}$, the first pulse heat electrons in the conduction band of TCO and moves them to the higher energy level while the second pulse deliver an additional energy to the electrons. Thus, the electrons in the conduction band moves to higher energy level - there is not enough time for electrons to decay and return to its initial state by the time the second pulse arrives to TCO. In a contrary, if separation between two pulses is longer than the thermalization time of electrons, $t_s>\tau_{th}$, the electrons return to its initial state before the second pulse arrives.

By combining both spatial and temporal summation we can bring a network closer to a real brain where neurons communicate with other neurons via action potential, *i.e.*, spikes. Network that operates in both spatial and temporal domains represent the third generation of neural networks – spiking neural networks (SNNs) [**5, 13, 45**]. Each neuron in such a network either generates complex time-resolved outputs in response to spatio-temporal inputs from upstream neurons or remain quiescent. Thus, it has already an in-built temporal short memory at the level of individual neurons.

**Electrical switching**

The permittivity of TCOs can be tuned either by the optical pulses through either generating or exciting electrons in the conduction band (interband or intraband absorption), so-called the all-optical switching mechanism, or by applying an electric field to the TCO, so-called the electrical switching mechanism. Under an applied voltage to the TCO thin layer, the carriers accumulate in TCO what induce the change in carrier concentration in TCO and, in consequence, reduces the permittivity in the accumulation region of TCO.

For a certain value of the incoming (input) power coming to the waveguide such that straight line (red or blue) in **Fig. 7a** crosses the black bell-shape curve, that defines the mode power attenuation of the waveguide for different plasma frequencies, at three points the bistability can be observed. Let's assume the input power of 300 mW coupled to the waveguide. For a low carrier concentration, the mode attenuation changes linearly with the carrier concentration. For a carrier concentration of $9.8 \cdot 10^{26}$ m$^{-3}$ the bistability appears defined by two stable attenuation points at 2.5 µm$^{-1}$ and 5.5 µm$^{-1}$. Here, the bistability is defined by the carrier concentration range from $9.8 \cdot 10^{26}$ m$^{-3}$ to $1.1 \cdot 10^{27}$ m$^{-3}$. At carrier concentration of $1.1 \cdot 10^{27}$ m$^{-3}$ the mode can be in two different attenuations points of 1.6 µm$^{-1}$ and 5.2 µm$^{-1}$ (**Fig. 7b**).

To show working principles of such a device let's assume a 300 mW of optical power coming to a device at which the ITO, one of the TCO materials, is initially at carrier concentration of, for example, $1.0 \cdot 10^{27}$ m$^{-3}$. By applying a short voltage signal to the ITO layer, the carrier concentration slightly increases to $1.025 \cdot 10^{27}$ m$^{-3}$. Thus, a mode power attenuation decreases very fast from 5.6 µm$^{-1}$ to 1.4 µm$^{-1}$. When the electrical signal is off, the mode power attenuation follows now the blue curve that defines the second stable point and reach mode power attenuation of 1.8 µm$^{-1}$ for the initial carrier concentration of $1.0 \cdot 10^{27}$ m$^{-3}$. In this way, under a short electrical signal the mode power attenuation drops from 5.6 µm$^{-1}$ to 1.8 µm$^{-1}$ what causes the change of the output optical power from a waveguide from 170 mW to 250 mW for a device length of only 100 nm, and from 18 mW to 120 mW for a 500 nm long device (**Fig. 7c**).

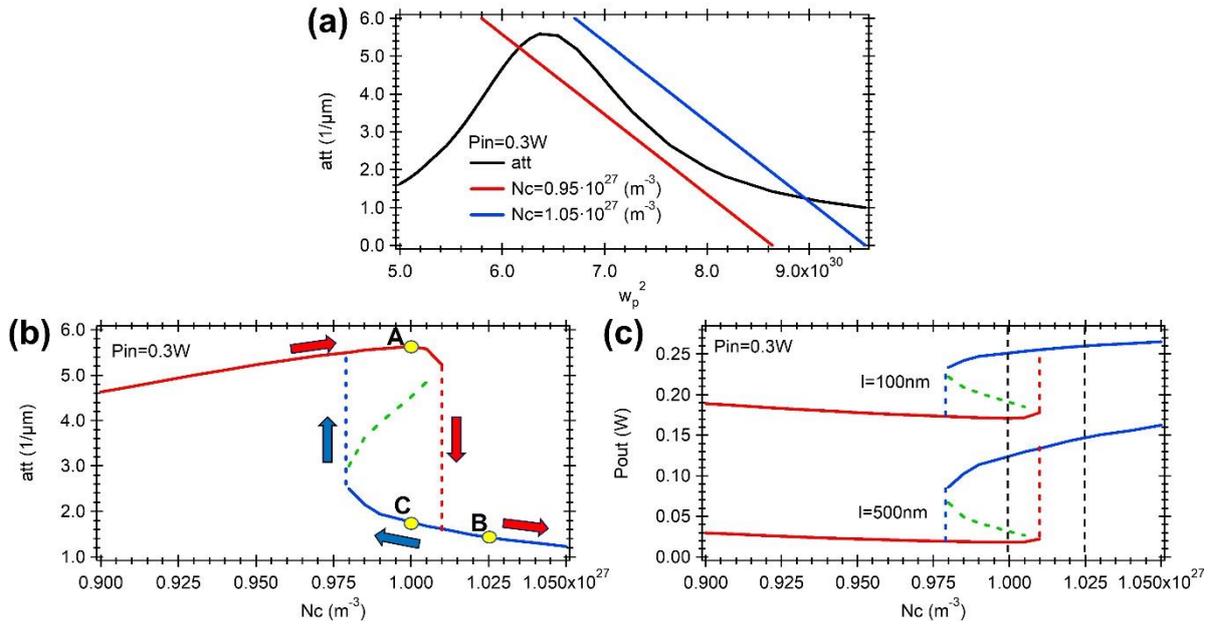

**Figure 7.** (a) Illustration of bistability and switching as plasma frequency changes in the waveguide for different carrier concentration in ITO. (b) Operation principle and (c) output power change from a device as a function of carrier concentration in ITO for different length of a device, l=100 nm and l=500 nm.

The proposed device can serve as a bistable non-volatile photonic memory that retain their values without active electrical power consumption after being set, *i.e.*, it survives after the voltage is turned off. The information can be reversed either by decreasing the carrier concentration below $9.8 \cdot 10^{26}$ m$^{-3}$ or by temporary turning off the optical power.

**Fig. 5 and 7** show a difference between an optical switching and electrical switching mechanisms. Under an applied voltage the carriers are excited from the valence band to the conduction band. These additional carriers lower the refractive index of the TCO via Drude dispersion making it more metallic and thus reducing the transmission. On the other hand, optical pulse with the energy lower than the bandgap heats the existing electrons in the conduction band towards higher energies. Due to the non-parabolic nature of the conduction band, the photoexcited electrons have a greater effective mass what decreases the frequency of the TCO, makes it more dielectric and, in consequence, increases the transmission. Thus, depending on the requirements, either all-optical or electrical switching mechanism can be involved providing a lot of flexibility to the system and network. Under an optical power coupled to a device the effective mass increases as the optical power increases what decreases the plasma frequency. In the contrary, as an applied voltage increases more carriers are accumulated in the TCO what causes the plasma frequency increases.

Furthermore, a voltage applied to TCO can serve as an excitatory or inhibitory inputs that depolarize or hyperpolarize the membrane potential in a brain – it can lower or increase the firing threshold of the neuron through a decreasing or increasing the carrier concentration into TCO [**48**].

**Conclusion**

In this work, we have examined a novel bistable device based on TCO materials arranged in a slot waveguide for application in artificial neural networks. Due to high nonlinearity and significant bistability combined with a small footprint and high switching speed the proposed device may turn out to be a game changer for photonic neural networks. It can operate in dual-mode, electrical and optical, what provides a lot of freedom in terms of design. Furthermore, it can combine both spatial and temporal summation and thus, in-built temporal short-memory, what can bring a network closer to a real brain and open a door for the third generation of neural networks – spiking neural networks.

**Acknowledgements**

J.G. thanks the "ENSEMBLE3 - Centre of Excellence for Nanophotonics, advanced materials and novel crystal growth-based technologies" project (GA No. MAB/2020/14) carried out within the International Research Agendas program of the Foundation for Polish Science co-financed by the European Union under the European Regional Development Fund and the European Union's Horizon 2020 research and innovation program Teaming for Excellence (Grant Agreement No. 857543) for support of this work.